\documentclass[prl,twocolumn,secnumarabic,tightenlines,aps,showpacs]{revtex4}
\usepackage{graphicx}
\begin{document}
\title{Conductivity of the defectless Graphene}
\author{A. Kashuba}
\affiliation{Max Planck Institute for Physics of Complex Systems, Noethnitzer str. 38, 01187 Dresden Germany}
\affiliation{Bogolyubov Institute for Theoretical Physics, 14-b Metrolohichna str., Kiev 03680 Ukraine}

\begin{abstract} Conductivity of the defectless, perfect crystal graphene is found at the neutrality point at zero temperature and in the limit of large dielectric constant of the substrate. The steady state of the graphene with weak current is assumed to be an ideal, rare plasma of particle and hole excitations governed by the Boltzmann kinetic equation.
\end{abstract}


\maketitle

The conductivity of a single-atom graphene layer as a function of carrier doping concentration shows a pronounced minimum at the neutrality, compensation point \cite{novos,tan,moroz}. Here both particle and hole excitations are present in the vicinity of two special points in the Brillouin zone where the dispersion becomes almost relativistic, massless, cone-shaped one \cite{w,geim}. In this case the theory predicts a universal conductivity for the non-interacting particles moving in a static disorder potential \cite{LFSG}. However, in graphene the experimental conductivity exceeds by few times the theoretical universal conductivity. Recent efforts \cite{devel} to clear this discrepancy mainly focus on the effects of disorder and have successfully explained a linear dependence of the conductivity as function of the doping concentration \cite{moroz}. Yet one point is being missed. Namely, the current of mobile charge excitations in a defectless, perfect crystal graphene can be changed, relaxed in the course of their mutual Coulomb interaction. And a question what is the conductivity of the defectless graphene is meaningful. Unlike the usual electron liquid where the current is synonymous in many situations to the momentum and, thus, where there is a rigorous current conservation by the Coulomb interaction. In this paper, the conductivity of the defectless graphene is found at the compensation point and in the limit of weak Coulomb interaction constant.

In the tight binding model on the honeycomb lattice, that may represent the band structure of the graphene, the current operator of few electrons: $i\sum (\psi^\dagger_\mathbf{x} \psi_{\mathbf{x}+\mathbf{a}}- \psi^\dagger_{\mathbf{x}+\mathbf{a}} \psi_\mathbf{x})$, does not commute with the Coulomb interaction operator. However, close to the cone apexes in the momentum space, or equivalently in the long wavelength limit, the current conservation is approximately restored. Nevertheless, this symmetry can not be projected onto the graphene directly as the latter has two types of excitations: particles and holes, in two relevant crystal bands. One band is completely filled and the other is empty, at the compensation point, and these two bands touch each other at the two cone apexes. Thus, the current operator has two differently ordered particle and hole terms: $\hat{\vec{j}}=\!\! \sum \psi^\dagger_{+}(\mathbf{p}) \partial\epsilon_+/\partial \mathbf{p} \psi_{+}(\mathbf{p}) - \psi_{-}(\mathbf{p}) \partial\epsilon_-/ \partial \mathbf{p} \psi^\dagger_{-}(\mathbf{p})$, as well as the inter-band current known as the Zitterbewegung. Here, $\mathbf{p}$ counts momenta in the Brillouin zone. To show how a commutation relationship changes when one half of the states is being filled, consider two operators: $\hat{A}= \psi^+A\psi= \psi^+_iA_{ij}\psi_j$ and $\hat{B}= \psi^+B\psi$, acting in the Fock space of electron states numerated by $i$. If the matrices $A$ and $B$ do commute: $[A,B]_-=0$ then the operators also commute: $[\hat{A},\hat{B}]_-= \psi^+[A,B]_-\psi=0$. However, if a subset of the states $\{i\}$ is filled then electron operators have to be normally ordered to represent excitations. In general, the ordered operators no longer commute: $[:\hat{A}:,:\hat{B}:]_-= \mathrm{Tr}((ANB-BNA)(1-N))$, where $N$ is the diagonal occupation matrix with the entries one for the filled subset $\{i\}$ and zero otherwise.

The main precondition for the non-conservation of the current is being close to the compensation point. A neutral cloud of non-interacting particles and holes responds to the electric field by separation: particles to one side whereas holes to the opposite side. In the momentum space, though, they all move in one direction. On the other hand, if a neutral particle-hole cloud is coupled by strong Coulomb forces it behave like a collection of pairwise neutral atoms. The response of these to the electric field is, initially, a polarization rather than a current. Therefore, a precise value of the conductivity is determined by the mutual Coulomb interactions. A microscopic process that changes the current is shown in the Fig.1. A particle 1 and a hole 2, considered as a pair, has zero net momentum and non-zero net current. The velocity of an electron in the state 2 is opposite to the velocity of an electron in the state 1. But a hole is the absence of an electron, therefore, the total current of the pair (1,2) is non zero. In the process of Coulomb interaction the pair (1,2) can be scattered into the new position (3,4) with the same total momentum and energy. We observe that the net current of this pair in the new state (3,4) has been reversed. Last argument is that although the kinetic energy in the graphene is artificially Lorentz invariant the Coulomb interaction is instantaneous and does violate this Lorentz invariance.

\begin{figure}
\includegraphics[width=7cm]{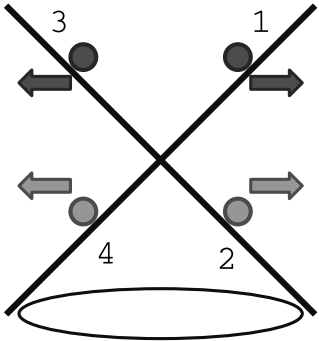}
\caption{\label{solut} A microscopic scattering process of the particle-hole pair (1,2) into the particle-hole pair (3,4) that conserves the momentum and the energy but changes the current. The excitation velocities are shown by the arrows. X-axis represent the momentum, whereas y-axis is the quasiparticle energy.}
\end{figure}

We study infinitely large, perfect 2D graphene layer on top of a dielectric substrate at zero temperature. Application of an electric field will inevitably create particle and hole excitation due to the Schwinger mechanism. The work of the electric field on these excitations will produce the Joule heat that will be transformed into the lattice vibrations near the graphene layer and will eventually escape into the bulk. In the balance, a steady distribution of particle and hole excitations will be established. We assume this state of the graphene to be an ideal, rare plasma with the excitation distribution given by the Fermi-Dirac function for some effective temperature $T^*$. The Hamiltonian in the long wavelength limit consists of the crystal band part \cite{geim}:
\begin{equation}\label{cbham}
\hat{H}=c\sum_{\mathbf{p}} \hat{\tau}^z \hat{\alpha}^x p_x+\hat{\alpha}_y p_y, 
\end{equation}
where $\hat{\alpha}^x,\hat{\alpha}^y,\hat{\tau}^z$ are the Pauli matrices [the first two act in the representation space of the crystal point group whereas the last one acts in the valley space] and the Coulomb interaction part. $c=1.1*10^8$ cm/s \cite{geim} is the characteristic band velocity of the graphene that determines the cone angle. The one-particle Hamiltonian (\ref{cbham}) can be diagonalize by the unitary transformation:
\begin{equation}
U=\frac{1+\hat{\alpha}^y}{\sqrt{2}} \exp\left( i\hat{\tau}^z \hat{\alpha}^z \frac{\phi}{2}\right)
\end{equation}
into the two crystal bands [two halves of the cone]: $\epsilon_{\tau\sigma}(\mathbf{p})=\tau\alpha |\mathbf{p}|$, where $\mathbf{p}=(p_x,p_y)$ and $\tau,\alpha=\pm 1$ are eigenvalues. At the compensation point the electronic state of graphene is determined by a dimensionless Coulomb coupling:
\begin{equation}\label{coupl}
g=\frac{e^2}{\kappa\hbar c}
\end{equation}
where $\kappa$ is the half of the dielectric constants of the substrate and the vacuum. For the graphene on top of Si substrate $g\approx 0.35$, whereas for the graphene on top of SiO$_2$ substrate $g\approx 0.8$. Coulomb interaction modify the crystal dispersion of the quasiparticle excitations \cite{geim}:
\begin{equation}\label{disp}
\epsilon(\mathbf{p})=c|\mathbf{p}|\left(1+\frac{g}{4}\log\frac{Q}{|\mathbf{p}|} \right)
\end{equation}
in the long wavelength limit $|\mathbf{p}|\ll Q$, where $Q$ is the Brillouin zone size, and this non-linearity of the dispersion will be important below.

In addition to the true non-equilibrium state of the graphene with the current described by the electron distribution function in the momentum space: $F_\alpha(\mathbf{p})$, where $\alpha=\pm 1$ specifies the two crystal bands, upper/lower halves of the cone, we consider also an imaginary 'equilibrium' state with relaxed, zero current but with the same excitation energy. We do not consider here the graphene states that have a particle-hole coherence of any kind [see e.g. \cite{ak}], as this may lead to the time dependence of the coherence order parameter averaging out its effect. In our solution $F_\alpha(\mathbf{p})$ does not depend on either spin or valley indices. The spin and valley spaces give the total degeneracy of electron states in the graphene $N=4$. As the total momentum of the scattered electrons is conserved in crystal [neglecting the Umklapp processes], we search for the graphene state with zero total momentum. In this state there are on average as many holes as particles in every small momentum cell. Therefore, we can describe the particles by the distribution function in the momentum space $F_+(\mathbf{p})$ whereas the distribution function for the holes $1-F_-(\mathbf{p})$ has to be the same. Thus, the distribution function possesses the particle-hole symmetry: $1-F_\alpha(\mathbf{p})=F_{-\alpha}(\mathbf{p})$. In the 'equilibrium' state of the graphene the electron distribution function:
\begin{equation}\label{zerodist}
f_\alpha(\mathbf{p})=\frac{1}{\exp(\alpha|\mathbf{p}|/\langle p\rangle)+1}
\end{equation}
makes the collision integral due to the Coulomb interaction to vanish for any scale parameter in the momentum space: $\langle p\rangle$. This scale also defines the effective temperature of the electrons in the 'equilibrium' state of the graphene: $T^*=c\langle p\rangle$. The distribution function Eq.(\ref{zerodist}) satisfies the electron-hole symmetry: $f_{-\alpha}(\mathbf{p})= 1-f_{\alpha}(\mathbf{p})$. From now on we will isotropically rescale the momentum space in the vicinity of the cone points to set $\langle p\rangle=1$.

The Boltzmann kinetic equation defines the steady distribution function in the state $\mathbf{p}$ balancing the two processes - the drift of excitations in the electric field and the redistribution of excitations during their collision:
\begin{equation}\label{BKE}
e\vec{E}\frac{\partial F_\alpha}{\partial \vec{p}}= St_{\alpha}(\mathbf{p})
\end{equation}
In the lowest order of the Coulomb coupling $g$, the second order Fermi golden rule, the collision integral reads [for short notations $\alpha_3,\alpha_4$ of the out-going electrons are inverted]:
\begin{eqnarray}\label{Stoss}
St_{\alpha_1}(\mathbf{p}_1)=\!\!\!\!\!\sum_{\alpha_2\alpha_3\alpha_4} \!\int\!\!\! \int\!\!\! \int \mathrm{Tr}_{\tau}\left( |V_{\alpha_1\alpha_2}^{\alpha_3\alpha_4} (\mathbf{p}_1\mathbf{p}_2 \mathbf{p}_3\mathbf{p}_4)|^2 \right) \times  \nonumber\\ (2\pi)\delta(\sum_{i=1}^4 \alpha_i\epsilon(\mathbf{p}_i))\left(\prod_{i=1}^4  F_{\alpha_i}(\mathbf{p}_i) - \prod_{i=1}^4  F_{-\alpha_i}(\mathbf{p}_i) \right)\times \nonumber\\  (2\pi)^2\delta(\mathbf{p}_1+ \mathbf{p}_2- \mathbf{p}_3- \mathbf{p}_4)\  \frac{d^2\mathbf{p}_2 d^2\mathbf{p}_3 d^2\mathbf{p}_4}{(2\pi)^6}
\end{eqnarray}
Below we use interchangeably the notation: $\mathbf{p}_1=\mathbf{p}$, $\mathbf{p}_2=\mathbf{p}'$, $\mathbf{p}_3=\mathbf{p}'+\mathbf{q}$ and $\mathbf{p}_4=\mathbf{p}'-\mathbf{q}$. The Coulomb matrix element is weakly screened, as the plasma of excitations is assumed to be rare in the limit of small $\langle p\rangle$ [with the Debye screening radius being large $R_D\sim \hbar^2c^2/e^2T^*$],:
\begin{eqnarray}\label{CoulMatr}
V_{\alpha_1\alpha_2}^{\alpha_3\alpha_4} (\mathbf{p}_1\mathbf{p}_2 \mathbf{p}_3\mathbf{p}_4)= \nonumber\\ \frac{1}{2}\left(\frac{2\pi e^2}{\kappa |\mathbf{p}_1-\mathbf{p}_3|} \frac{1-z_1z_3^*}{2} \frac{1-z_2z_4^*}{2} \ \delta_{\tau_1\tau_3}\delta_{\tau_2\tau_4} \right. \nonumber\\ \left. - \frac{2\pi e^2}{\kappa |\mathbf{p}_2-\mathbf{p}_3|} \frac{1-z_2z_3^*}{2} \frac{1- z_1z_4^*}{2} \ \delta_{\tau_1\tau_4}\delta_{\tau_2\tau_3} \right)
\end{eqnarray}
where the notation $z_i=\alpha_i(p^x_i+ip_i^y)/|\mathbf{p}_i|$ is used. $\tau$ indices run over the total $N=4$ spin-valley degeneracy space of the graphene. The square of the matrix element Eq.(\ref{CoulMatr}) includes two terms - the direct and exchange ones. The exchange term vanishes when two scattering excitations have different spins or valleys.

In the graphene state with current the electron distribution function can be written as:
\begin{equation}\label{nonEqDis}
F_\alpha(\mathbf{p})=\frac{1}{\displaystyle \exp\left[ \alpha |\mathbf{p}|+ \alpha (e\vec{E}\cdot\vec{p})\ \chi(|\mathbf{p}|)/ |\mathbf{p}| \right] +1}
\end{equation}
where $\chi(\mathbf{p})$ defines the perturbation of the distribution in electric field. It is better to satisfy the condition $\chi(\mathbf{p})\to 0$ as $|\mathbf{p}|\to 0$. Also, $\chi(\mathbf{p})$ in Eq.(\ref{nonEqDis}) explicitely conserves the number of electrons, their total energy and their total momentum. We linearize the Boltzmann kinetic equation Eq.(\ref{BKE}) with respect to $\chi(\mathbf{p})$. The linearized collision integral becomes a matrix, which is symmetric due to a detailed balance of the direct and time-reversed processes in the steady state.  The current in the linear responce reads [the band velocity $\vec{v}_\alpha=\alpha \vec{p}/|\mathbf{p}|$]:
\begin{equation}\label{current}
\vec{j}[\chi]=-N\frac{e^2}{\hbar} \sum_{\alpha} \int \frac{\chi(|\mathbf{p}|)}{|\mathbf{p}|^2} \vec{p}(\vec{p}\cdot \vec{E}) f_\alpha(\mathbf{p}) f_{-\alpha} (\mathbf{p})\ \frac{d^2\mathbf{p}}{(2\pi)^2}
\end{equation}
For strictly linear dispersion and for the collinear orientation of all momenta $\mathbf{p}||\mathbf{p}'||\mathbf{q}$ the argument of the energy delta-function in Eq.(\ref{Stoss}) becomes degenerate, i.e $\alpha_1 |\mathbf{p}|+ \alpha_2|\mathbf{p}'|+ \alpha_3|\mathbf{p}+\mathbf{q}| +\alpha_4|\mathbf{p}'-\mathbf{q}|=0$ for any $|\mathbf{p}'|$ and $|\mathbf{q}|$ provided three conditions are met: $\alpha_1= \alpha_2\mathrm{sgn}(\mathbf{p}\cdot\mathbf{p}_2)= -\alpha_3 \mathrm{sgn}(\mathbf{p} \cdot\mathbf{p}_3) =-\alpha_4 \mathrm{sgn}(\mathbf{p} \cdot\mathbf{p}_4)$. Taking the direction of vector $\mathbf{p}$ as $x$ and expanding around the collinear configuration of four momenta we find:
\begin{equation}\label{dE}
\Delta E= \frac{{p'}_{y}^{2}}{2p'}- \frac{q_y^2}{2(p+q)}-\frac{(p'_{y}-q_y)^2}{2(p'-q)}+ \frac{g}{4}\sum_i p_i\log|p_i|
\end{equation}
where $p_i=p_{ix}$ and where the effect of the renormalization of the velocity $\delta c=(g/4)\log(Q/\langle p\rangle)$ is omitted. The last term in Eq.(\ref{dE}) can be approximated as $\pm g$. The integration of the energy delta-function $\delta(\Delta E)$ with respect to the $y$ momentum components gives the Jacobian $\sqrt{pp'(p+q)(p'-q)}/|p|$, provided $pp'(p+q)(p'-q)>0$, and the large logarithm $2\log(1/g)$. Thus, in this leading large logarithm approximation:
\begin{equation}\label{largelog}
\frac{1}{2\pi}\ \log\left(\frac{1}{g}\right) \gg 1
\end{equation}
the linearized Boltzmann kinetic equations reads:
\begin{eqnarray}\label{reduceBKE}
\lambda\int\!\!\!\!\int_{-\infty}^{+\infty}\!\!\!\! \frac{\chi(|p|)+ \chi(|p'|) - \chi(|p+q|)-\chi(|p'-q|)}{(e^{\displaystyle p}+1) (e^{\displaystyle p'}+1) (e^{\displaystyle -p-q}+1) (e^{\displaystyle -p'+q}+1)} \nonumber\\ \times\frac{\sqrt{pp'(p+q)(p'-q)}}{q^2} \frac{dp'dq}{2\pi}= \frac{-|p|}{(e^{\displaystyle p}+1)(e^{\displaystyle -p}+1)}
\end{eqnarray}
where $\lambda=2Ng^2\log(1/g)$ is the Coulomb integral, and the condition $pp'(p+q)(p'-q)>0$ is enforced in the integrand. The exchange term vanishes in the leading logarithm approximation. The Debye screening mass makes the integral in Eq.(\ref{reduceBKE}) to converge as the principle value in the vicinity of $q=0$. Due to few symmetries of the integral in Eq.(\ref{reduceBKE}): $(p\leftrightarrow p'\ ,\ q\leftrightarrow -q)$, $p\leftrightarrow -p-q$ and $p'\leftrightarrow -p'+q$, the Eq.(\ref{reduceBKE}) is a symmetric operator. Thus, the Boltzmann kinetic equation is the variation of the functional: $\mathcal{R}[\chi]-\Sigma[\chi]$, where
\begin{eqnarray}
\mathcal{R}[\chi]=\frac{\lambda}{8}\int\!\!\!\!\int\!\!\!\!\int_{-\infty}^{+\infty} \frac{\sqrt{pp'(p+q)(p'-q)}}{q^2} \frac{dpdp'dq}{(2\pi)^2} \nonumber\\ \times \frac{\left(\chi(|p|)+ \chi(|p'|)- \chi(|p+q|)- \chi(|p'-q|)\right)^2}{(e^{\displaystyle p}+1) (e^{\displaystyle p'}+1) (e^{\displaystyle -p-q}+1) (e^{\displaystyle -p'+q}+1)} \nonumber\\ \Sigma[\chi]=-\int\frac{p\chi(p)}{(e^{\displaystyle  p}+1)(e^{\displaystyle -p}+1)} \frac{dp}{2\pi}
\end{eqnarray}
The existence of this positively defined functional $\mathcal{R}[\chi]$ proves that the conductivity is positive. Indeed, in the minimum: $\mathcal{R}[\chi]- \Sigma[\chi]<0$ because $\mathcal{R}[0]-\Sigma[0]=0$. The conductivity $\sigma=\Sigma[\chi]>\mathcal{R}[\chi]>0$.

The equation (\ref{reduceBKE}) is contradictory and has no solution as the integral over all $p$ applied to the left hand side is zero. It means that the leading, large logarithm approximation is insufficient. However, from the all next order terms of the Boltzmann kinetic equation we need only their combined action on the homogeneous, in the momentum space, mode $\chi(\mathbf{p})=\chi_0=const$, which is being neglected by the leading term Eq.(\ref{reduceBKE}). The 'leading' order of this 'subleading' term is $g^2$ without the large logarithm. The mode $\chi_0$ arises in the process of parallel shift of all momenta $|\mathbf{p}+\mathbf{a}|= \mathbf{p}+(\mathbf{p}\cdot \mathbf{a})/|\mathbf{p}|$. We write the 'subleading' term as a projection: $Ng^2 |\Phi(\mathbf{p}) \rangle \langle \Phi(\mathbf{p})|$, onto a function $\Phi(\mathbf{p})$. The function $\Phi(\mathbf{p})$ can be found in the closed form as an integral. We parameterize the four momenta of two scattering electrons by $p_i=\alpha_i|\mathbf{p}_i|$. These also define the mutual angles of the momenta upto to the four-fold discreet flip transformations. The collision integral gives the function $\Phi(\mathbf{p})=St_\alpha(\mathbf{p})[\chi=1]$, independent of $\alpha$, and we set $\alpha=+1$. Its exchange part is rather complicated whereas the direct part reads:
\begin{eqnarray}\label{Phi}
\Phi_d(p)=\int\!\!\!\!\!\int_{-\infty}^{+\infty} \frac{1}{(e^{\displaystyle p}+1) (e^{\displaystyle p_2}+1) (e^{\displaystyle p_3}+1) (e^{\displaystyle p_4}+1)} \nonumber\\ \left(\sqrt{us}- \left(2Q^2-u-s\right) \mathrm{Arcth} \sqrt{\frac{u}{s}} + \right. \nonumber\\ \left. +2\sqrt{Q^2-u}\sqrt{Q^2-s}\ \mathrm{Arcth} \sqrt{\frac{(Q^2-u)s}{(Q^2-s)u}} \right) \frac{dp_2 dp_3}{2\pi\ u}
\end{eqnarray}
where $p_4=-p-p_2-p_3$ due to the energy conservation, and the parameterization: $Q=p+p_3=-p_2-p_4$, $u=4pp_3$, $s=4p_2p_4$ [with $Q^2>|u|,|s|$] is used, satisfying the condition $us>0$ in the Eq.(\ref{Phi}). As long as $\Phi(p)$ is known and the homogeneous mode $\chi_0$ is singled out:  $\chi(\mathbf{p})=\chi_0+\chi_1(\mathbf{p})$, the Boltzmann kinetic equation reads:
\begin{equation}\label{BKEextra}
Ng^2\Phi(p)\chi_0+ \frac{\delta \mathcal{R}}{\delta\chi}[\chi_1] =-\frac{|p|}{(e^{\displaystyle p}+1)(e^{\displaystyle -p}+1)}
\end{equation}
In the large logarithm limit Eq.(\ref{largelog}), the function $\chi_1(\mathbf{p})$ is relatively small $|\chi_1(\mathbf{p})|\ll |\chi_0|$. Integrating Eq.(\ref{BKEextra}) with respect to $p$ eliminates all non-homogeneous modes in the kinetic equation that determines $\chi_1$ and leaves the equation for $\chi_0$ only:
\begin{equation}
Ng^2C\chi_0= \int_0^{+\infty}\!\!\!\! \frac{-|p|\ dp}{(e^{\displaystyle p}+1)(e^{\displaystyle -p}+1)} =-\log(2)
\end{equation}
Solution is: $\chi_0=-\log(2)/NCg^2$, where $C$ is the average of the collision integral over the homogeneous mode:
\begin{equation}\label{C}
C=\int_0^{+\infty}\!\!\!\Phi(p) dp
\end{equation}
We estimate numerically the direct $C_d\approx 0.69$ and the exchange $C_{ex}\approx -0.1/N$ parts of $C=C_d+C_{ex}$. Thus, the distribution of excitations with current is approximately the same as without current just translated in parallel in the momentum space by a vector proportional to the electric field. This solution, though, does not vanish in the momentum origin. We estimate numerically that adding some gap to the collision integral in the origin $\Delta \delta(p) \delta(p')$ gives a necessary crossover of $\chi(\mathbf{p})$ to zero on the scale $|\mathbf{p}|<g$, and as $g\to 0$ this feature has negligible effect.

The conductivity is determined by the current Eq.(\ref{current}): $\vec{j}[\chi]= -N\log(2) (e^2/\hbar)\ \chi_0\ \vec{E}$, neglecting small $\chi_1$ contribution. From this equation, we finally obtain the conductivity of the defectless, perfect crystal graphene:
\begin{equation}\label{result}
\sigma=\frac{e^2}{\hbar}\frac{\log^2(2)/C}{2\pi g^2}
\end{equation}
in the limits of large dielectric constant of the substrate $g\to 0$ and, also, the large logarithm  $\log(1/g)/(2\pi)\to \infty$. Remarkably, the conductivity Eq.(\ref{result}) does not depend on the 'heating' scale $\langle p\rangle=T^*/c$. Inclusion of specific mechanisms of energy relaxation due to say electron-phonon interaction is necessary to determine this scale. At any rate, $T^*$ is small in the limit of weak electric field $T^*\sim E^{\gamma}$. However, the power $\gamma$ can be rather small in the realistic electron-phonon models. For the Coulomb coupling $g\approx 0.35$ the minimum conductivity in the Eq.(\ref{result}) corresponds to the experimental values around $\rho_{max}\approx 4$ kOhm. Our numerical estimation of the different parts of the Boltzmann kinetic equation shows that the large logarithm approximation begins to work at around $g<0.2$ whereas for the experimental condition $g\sim 0.4$ an increase of the conductivity Eq.(\ref{result}) by 30\% or so has to be expected.

As the gate voltage breaks the particle-hole symmetry and creates a net charge on the graphene: $e(N_h-N_e)$, the electric field induces non-conservation of the total momentum: $d\vec{P}/dt= e(N_h-N_e)\vec{E}$. This runaway evolution of the excitation distribution can not be controlled by their mutual Coulomb interactions alone, as they conserve the momentum, and some defects violating the translational symmetry is required to stabilize the steady state. This situation can be treated in a similar fashion.

\end{document}